# Faraday effect and fragmentation of ferromagnetic layers in multilayer Co/Cu(111) nanofilms

Irene N. Lukienko[a,*], Mykola F. Kharchenko[a], Alexey V. Fedorchenko[a], Ivan A. Kharlan[a], Olga P. Tutakina[a], Olexandr N. Stetsenko[b],
Cristina S. Neves[c], Andrei N. Salak[c]

[a]*Institute for Low Temperature Physics and Engineering, 47 Nauky Ave., 61103 Kharkiv, Ukraine*
[b]*National Technical University "Kharkov Polytechnical Institute", Frunze St. 21, 61002 Kharkiv, Ukraine*
[c]*Department of Materials and Ceramics Engineering and CICECO-Aveiro Institute of Materials University of Aveiro, 3810-193 Aveiro, Portugal*

With purpose to investigate influence of magnetically non-active metal layers on the Faraday effect in multilayer Ferromagnetic/Normal metal films, dependences of the Faraday rotation angles of the light polarization plane on magnetic field have been studied in multilayer [Co/Cu] nanofilms. It was revealed that the Faraday rotation $\varphi$ varies with thickness of the Cu layers $d_{Cu}$. This $\varphi(d_{Cu})$ dependence consists of the monotonic component, namely a gradual rise of the angle with increase of $d_{Cu}$, and the non-monotonic one represented by two minima. The monotonic changes of the Faraday rotation were satisfactory described in frames of the effective medium method. Two minima are explained with the Co layer's fragmentation due to influence of size electron quantization in the Cu layers on formation of Co clusters during deposition of the films.

**Keywords**: multilayer Co/Cu nanofilms, Faraday effect, superparamagnetic clusters, quantum size effect, fragmentation of nanoscale magnetic layers

## 1. Introduction

Multilayer epitaxial metallic nanofilms, which are widely used in different areas of information and sensor technology, continue to be promising objects for scientific study [1-4]. Among the "Ferromagnetic/Normal metal" (FM/NM) systems the Co/Cu periodic structures remain attractive due to their magnetoresistive properties and manufacturability [4,5]. The Giant Magnetoresistive effect (GMR) in such nanofilms is attributed to the antiferromagnetic (AF) exchange coupling between Co layers through conductivity electrons of copper and depends on both the thickness of Cu layers and structure of the Co/Cu interfaces [6-7].

Ferromagnetic Co layers of the multilayer Co/Cu nanofilms, prepared by different methods, often are non-continuous and have no sharp interfaces. The Co layers contain small FM clusters, which behave as superparamagnetic (SPM) particles [8-10]. Magnetoresistance of these nanofilms is caused by spin-dependent scattering of conductivity electrons on the SPM clusters [11-12]. Many questions in relation to connection between parameters of the SPM clusters, their quantity and technology of preparation of the nanofilms remain open so far [8-10, 13].

In our previous studies [14,15] of magnetoresistive properties and the magneto-optic longitudinal Kerr effect of the $[Co/Cu(111)]_m$ multilayers, obtained by the magnetron sputtering method, we found that the SPM clusters are smaller in the films, for which there is an antiferromagnetic exchange coupling between the Co layers. It was supposed that this feature is caused by the electronic quantum size effect in the Cu layers on formation of the Co layers during their deposition. The inhomogeneous electric field of electron standing waves in the Cu layer affects the deposition of Co atoms, and under these conditions a more fragmented cobalt layer containing smaller SPM clusters is formed.

Usually, due to high reflectivity and sufficiently large magneto-optical coefficients, the Kerr effects are used to research ferromagnetic metals. In a case of the Kerr effect the FM/NM interfaces play a major role in formation of magneto-optical response, but for the Faraday effect the response is formed by the whole thickness of the FM layers. This circumstance allows to use the Faraday effect as a tool to get additional information about FM layers in multilayer films having a thickness up to tens of nanometers, at which they are still sufficiently transparent. In this work we study the behavior of the Faraday effect in multilayer Co/Cu having different thickness of the Cu layers in order to determine mechanisms of influence of magnetically inactive metal layers on the Faraday effect in the magnetoresistive FM/NM nanofilms prepared by the magnetron sputtering.

## 2. Nanofilm structure and experimental methods

The multilayer $[Co(0.8\ nm)/Cu(d_{Cu})]_m$ nanofilms (Fig. 1) were obtained by the magnetron sputtering method in the vacuum setup, where the residual atmosphere was $10^{-6}$ Torr. The working pressure of argon during sputtering did not exceed $1.3 \cdot 10^{-3}$ Torr. Fluorphlogopite mica was used as a transparent substrate, which has optical properties of a weak birefractive biaxial crystal. The Cu buffer layer with thickness 5 nm was deposited on the mica substrate before deposition of the multilayer structure. The deposition rates for Co and Cu were 0.045 nm/s and 0.058 nm/s, respectively. Thicknesses of the layers were defined by the deposition time. Calibration of the deposition time was made by the method of multi-beam optical interferometry with error less than 2% [16].

The prepared films $[Co\ (0.8\ nm)\ /\ Cu\ (d_{Cu})]_m$ contained 20 cobalt layers and 19 cooper layers located between them. All cobalt layers had identical thickness 0.8 nm (4 atomic layers). Twelve multilayer nanofilms in which Cu layers had thicknesses $d_{Cu}$ = 0.6, 0.7, 0.9, 1.0, 1.1, 1.2, 1.35, 1.5, 1.7, 1.8, 1.9, and 2.0 nm were studied (Fig. 1). The top Cu layer was 1.25 nm thick. All films had a layered granular columnar structure, in which the granules were multilayer columns with transverse sizes of about 8 - 10 nm (Fig. 2).

The electron diffraction studies were performed using an EMV-100AK transmission electron microscope (TEM) operating with the accelerating voltage of 100 kV. It was found that the copper and cobalt layers had a face-centered cubic structure with the crystallographic planes of (111) Co and Cu oriented parallel to the substrate [16].

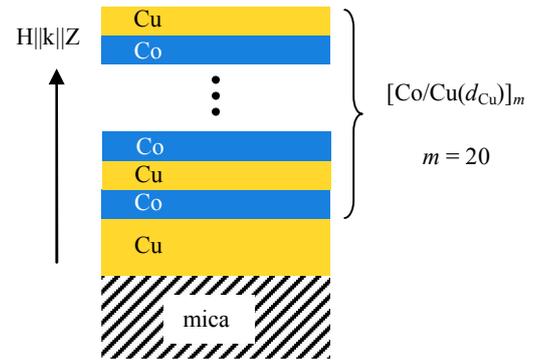

Fig. 1. Schematic view of the $[Co(0.8\ nm)/Cu(d_{Cu})]_m$ multilayer nanofilms.

For measurements of the Faraday effect, the polarization modulation method was used. YBi-iron-gallium garnet served as the working medium of the magneto-optic modulator. The compensator in which the angle of rotation of the light polarization plane was proportional to the current through the solenoid was used for calibration of the Faraday rotation angles. Optical glass was used as a working medium of the magneto-optic compensator. A helium-neon laser with radiation wavelength $\lambda$ = 632.8 nm served as a light source. The $[Co/Cu(d_{Cu})]_{20}$ films were oriented perpendicular to the magnetic field vector. The dependences of the Faraday rotation angle $\varphi$ on the magnetic field strength $H$ were measured.

During the measurements, an incident light propagated in the first instance through the mica substrate and then through the $[Co/Cu(d_{Cu})]_{20}$ film to minimize an influence of birefringence in mica. The samples were always oriented in such a way that polarization of the incident light was parallel to the plane of optical axes of the mica substrate.

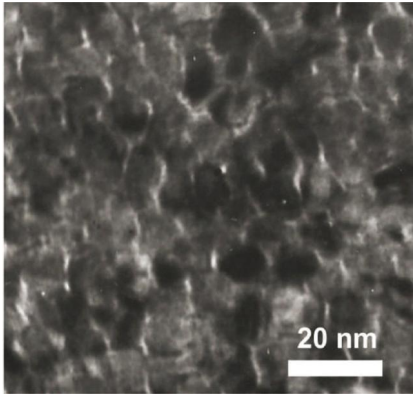

Fig. 2. TEM image of grains in [Co/Cu(1 nm)]$_{20}$.

Contribution to the Faraday rotation from the mica substrate did not exceed 3×10$^{-5}$ deg/kOe. Contribution to the total rotation of polarization plane from the Faraday effect in the lenses, caused by the stray field of electromagnet, was less than 2×10$^{-4}$ deg/kOe. These both contributions were measured and subtracted from the experimental $\varphi(H)$ dependences. Magnetization was measured for the nanofilms with $d_{Cu}$ = 0.7 and 0.9 nm at 300 K using the SQUID-magnetometer. Surface potential images of the films with $d_{Cu}$ = 0.9, 1.5, and 1.8 nm were obtained using a Multimode Atomic Force Microscope (Nanoscope IV from Veeco). The images were acquired in tapping mode, using silicon probes, operating at a resonance frequency of about 320 kHz and a force constant of 42 N/m. Image analysis was performed using Gwiddeon Software (version 2.37).

## 3. Results and discussion

The $\varphi(H)$ dependences without contributions from the lenses and the mica substrate are shown in Fig. 3 for several [Co/Cu($d_{Cu}$)]$_{20}$ films with the copper layer thickness $d_{Cu}$ = 0.6, 0.9, 1.35, 1.8, 1.9, and 2.0 nm. The $\varphi(H)$ dependences are the same upon both for increasing and decreasing of the magnetic field strength without hysteresis loops in the frames of scatter of the experimental points. Therefore, the magnetic state of the nanofilms was considered as equilibrium enough.

The magnitudes of $\varphi$ versus the Cu layer thickness are shown in Fig. 4. It should be noted that the angles of Faraday rotation at the same magnetic field strength are different for most of the films. This feature seemed to be unusual for the films under study, in which the total nominal thickness of the magneto-optical Co layers is invariable. The $\varphi(d_{Cu})$ dependences demonstrate two overlapping features, namely monotonic rise of Faraday rotation angle with increasing of the Cu layer thickness and two minima at certain Cu layer thickness values: near $d_{Cu}$ = 1.0 and 1.8 nm. The monotonic rise is shown in Fig. 4 as dashed lines.

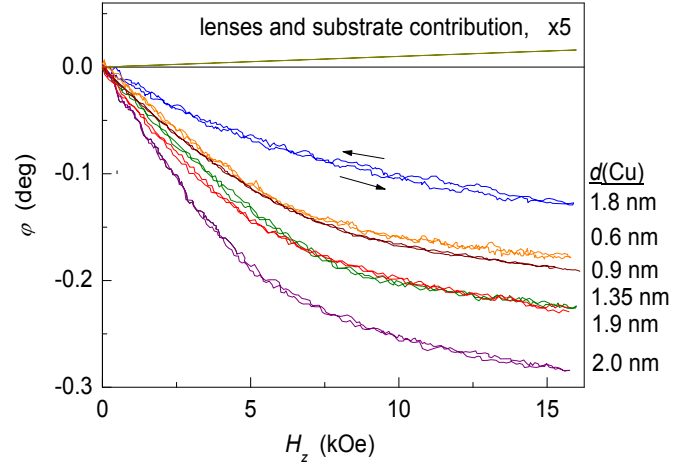

Fig. 3. The magnetic field dependences of Faraday rotation angle for several [Co/Cu($d_{Cu}$)]$_{20}$ films under study. Two curves for every film correspond to the opposite orientation of the applied magnetic field. The vertical axis corresponds to the absolute magnitudes of the Faraday rotation angles. The summary contribution from the lenses and the mica substrate is shown multiplied by factor of 5.

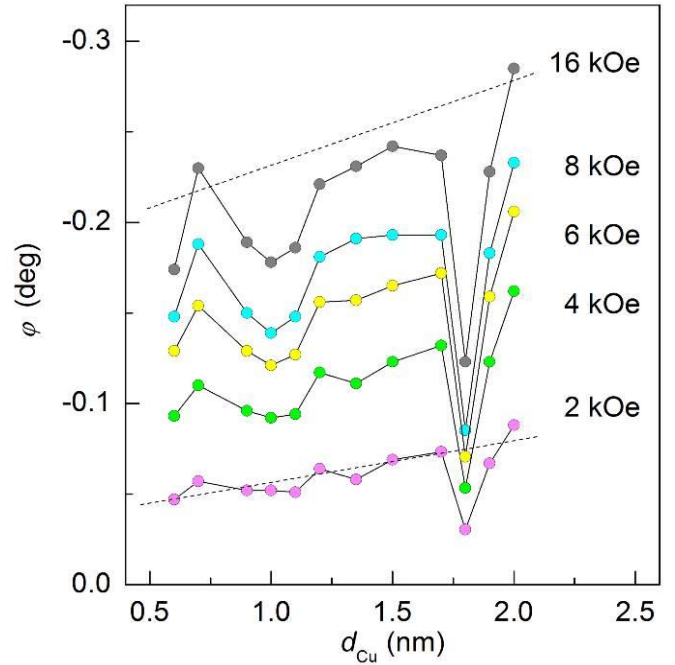

Fig. 4. The Faraday rotation angles in the [Co/Cu($d_{Cu}$)]$_{20}$ films at different strength of the applied magnetic field as a function of the Cu layer thickness. The dashed lines indicate a monotonic rise of the Faraday angles $\varphi$.

### 3.1 Monotonic variation of the Faraday rotation vs. $d_{Cu}$

It should be noted that the field dependences $\varphi(H)$ in the region of weak magnetic fields (less than 3 kOe) are linear ones indeed but are not such that are asymptotically approaching to linear ones. These field intervals of linearity were found for all the films by plotting the differences $\varphi - \beta H$ as a function of $H$. The $\beta$ coefficients were chosen in order to compensate the linear rise of $\varphi$ and to obtain the plateau on the dependences $\varphi - \beta H = f(H)$ for every film. Figure 5 illustrates the existence of the linearity intervals $2H^*$ in the $\varphi(H)$ dependences in particular nanofilms.

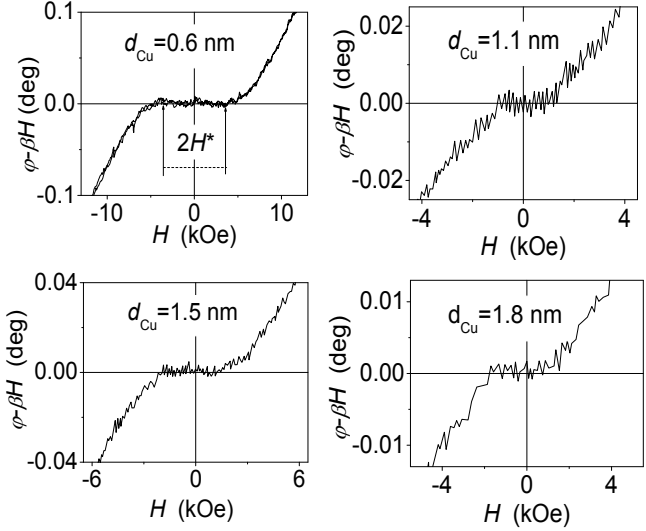

Fig. 5. Illustration of existence of the linearity intervals in the experimental $\varphi(H)$ dependences, which are transformed in horizontal plateaus in the magnetic field dependences of $\varphi-\beta H$.

Magnetic anisotropy of Co / Cu nanofilms is the "easy plane anisotropy" because the magneto-dipole contribution to anisotropy energy is predominant [17]. The observed non-asymptotic linear parts of the $\varphi(H)$ dependences

$$\varphi(H) = \beta H = V m d_{Co} H, \quad (1)$$

when magnetic field is perpendicular to the film plane, point to homogeneous rotation of magnetic moments in the Co layers to the field direction at $H<H^*$. This property allows to determine the magneto-optical coefficients of proportionality between the Faraday rotation and magnetization - the effective Kundt coefficients for all the films:

$$K = \frac{\varphi}{m d_{Co} M}. \quad (2)$$

In the case of homogeneous rotation of the magnetic moments of the ferromagnetic layers, the projection of the magnetization of the composite film on the field direction is determined by demagnetization factor of the film, $N$, as $M = \frac{1}{N} H$, and the Faraday rotation angle can be expressed in the following form:

$$\varphi = \frac{1}{N} K\, m d_{Co}. \quad (3)$$

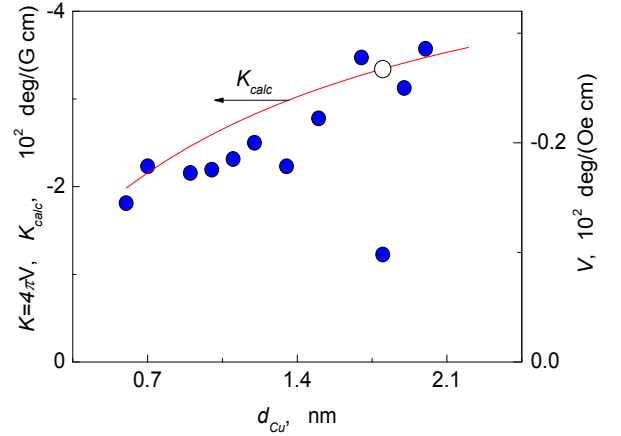

Fig. 6. The Cu layer thickness ($d_{Cu}$) dependences of the coefficient $V$ (solid circles, right scale) and $K=4\pi V$ (left axis), determined from the experiment at $H<H^*$ (Eq. (1), (4)), and the Kundt $K_{calc}$ coefficient, calculated according to Eq. (6) (solid line, left axis) in assumption that magnetization of the ferromagnetic layers $M_0$ is invariable. Meaning of the value denoted by open circle is explained in the text.

Using the $N$-factor and the obtained magneto-optical coefficient $V$ it is possible to perform the effective Kundt coefficient as

$$K = NV \quad (4)$$

The demagnetization factor $N$ of the multilayer in direction perpendicular to its plane is close to $4\pi$. If we consider the Co/Cu multilayer as a set of unlinked ferromagnetic disks whose planes coincide with the plane of the film and which are evenly distributed in the Cu matrix with effective density $\rho = d_{Co}/(d_{Co}+d_{Cu})$, the demagnetization factor of such a composite flat system can be estimated as $N = (1-\rho)N_{disk\parallel} + 4\pi\rho$, where $N_{disk\parallel} \approx 4\pi/(1+1.6\frac{d_{Co}}{D})$ for the disk diameter $D \gg d_{Co}$ [18,19]

The factor $N/4\pi$ varies from 0.90 to 0.95 with the thickness of the copper layers $d_{Cu}$ (from 0.6 to 2.0 nm) and with the disk diameters $D$ (from 8.0 to 10 nm). Taking into account that the Co disks of the adjacent multilayer grains form planes, $N$ should be even closer to $4\pi$. In order to plot the dependence of $K(d_{Cu})$, we assumed that the ferromagnetic layers in the columnar grains remain similar to disks for all films, and the coefficients of $N$ for them are the same and equal to $4\pi$. The error in the determination of $K$, which caused by the difference of the coefficient $N$ from $4\pi$, does not exceed 10%.

The values of the coefficients $V$ and $K=4\pi V$ depending on the thickness of the copper layers in the films are shown in Fig. 6. As seen from the Figure, the magneto-optical coefficient $V$, determined for weak fields, where homogeneous rotations of the magnetic moments of the Co layers occur, varies with the thickness of the Cu layers, in accordance with the Faraday rotation changes in stronger fields. The $V(d_{Cu})$ dependence is similar to $\varphi(d_{Cu})$ one and has monotonic and non-monotonic components.

The influence of NM layers on magneto-optical effects in FM / NM multilayer structures has been studied theoretically and experimentally in many papers [20-31]. Among different mechanisms of the influence, the interface hybridization of the electronic orbitals of the NM and FM atoms, which leads to changes of energy level structure of both metals, was considered. These changes can cause both increase and decrease of magneto-optical effects depending on the light wavelength and thickness of the metal layers [20]. As it has been reported in Ref. [21] the changes in spectra of the magneto-optical polar Kerr effect for the Co/Cu nanofilms due to hybridization do not exceed ten of percent. For the Faraday rotation influence of the interfaces is much weaker.

Spin polarization of conduction electrons of NM layers which are adjacent with FM layers can also affect the magnitudes of magneto-optical effects [22]. This mechanism depends on thickness of the NM layer. Changes of magneto-optical properties are results of formation of new resonance states (quantum well states) of spin-polarized conduction electrons in the NM layer, which induce the antiferromagnetic exchange coupling between the FM layers. Increasing of the magneto-optical Kerr rotation due to this quantum size effect does not exceed 10% [23, 24, 25].

The magnitudes of magneto-optical effects in dependence on thickness and number of the FM and NM layers are usually calculated using the transfer-matrix method [26-28]. However, in the case when period of the structure $\Lambda = d_{FM} + d_{NM}$ is much less than the light wave length, $\Lambda << \lambda$, magnitudes of magneto-optical effects can be estimated by the method of effective medium [29,30]. This method was used for consideration of dispersion of optical and magneto-optical parameters in two-component multilayer Co/Cu nanofilms in order to compare calculated and experimental data. The difference between the experimental and the calculated results in this method did not exceed 20% [31].

In the present study, it was supposed that the components of the dielectric permeability tensors $\varepsilon_{ij}$(Co) and $\varepsilon_{ij}$(Cu) are the same for all the [Co(0.8 nm)/Cu($d_{Cu}$)]$_{20}$ nanofilms. The angles $\varphi$ of Faraday rotation for the magnetized metal films can be written as [32]

$$\varphi = \frac{\pi d_{eff}}{\lambda} \frac{\text{Im}(N_{eff})\text{Re}(\varepsilon_{xy}^{eff}) - \text{Re}(N_{eff})\text{Im}(\varepsilon_{xy}^{eff})}{\text{Re}(N_{eff})^2 + \text{Im}(N_{eff})^2} \quad (5)$$

where $d_{eff} = m(d_{Co} + d_{Cu})$ is the thickness of the effective magneto-optical film, equal to the sum of the thicknesses of all magneto-opticaly active and non-active layers,

$$N_{eff} = \sqrt{\varepsilon_{xx}^{eff}} = \sqrt{\frac{\varepsilon_{xx}(Co)d_{Co} + \varepsilon_{xx}(Cu)d_{Cu}}{d_{Co} + d_{Cu}}}$$

is the effective refractive index, and

$$\varepsilon_{xy}^{eff} = \frac{\varepsilon_{xy}(Co)d_{Co} + \varepsilon_{xy}(Cu)d_{Cu}}{d_{Co} + d_{Cu}}$$

- the effective non-diagonal components of the dielectric permeability tensor. For copper, $\varepsilon_{xy}$(Cu) = 0.

Using Eq. (5), the magnitudes of $\varphi_{calc}$ and Kundt coefficient

$$K_{calc}(d_{Cu}) = \frac{\varphi_{calc}(d_{Cu})}{md_{Co}M_0} \quad (6)$$

were calculated for different $d_{Cu}$ values (solid line in Fig. 6).

The diagonal components of the dielectric permeability tensor were taken for copper as $\varepsilon_{xx}$(Cu) = -11.64-$i$1.64 from Ref. [33] and for cobalt as $\varepsilon_{xx}$(Co) = -11.5-$i$18.32 from Ref. [34]. The non-diagonal component was fitted as $\varepsilon_{xy}$(Co) = 0.36-$i$0.057. It should be noted that the diagonal $\varepsilon_{xx}$ components and, especially, the non-diagonal $\varepsilon_{xy}$ components or magneto-optic coefficients $Q = Q_1 + iQ_2 = i\varepsilon_{xy}/\varepsilon_{xx}$ reported by different authors (see Table 1) are considerably different.

Magnetization of the FM subsystem, $M_0$, in saturated state was taken as 495 G for all the films. This value was obtained for the film with $d_{Cu}$ = 0.7 nm from the SQUID-magnetometer measurements and chosen because considerable part (not less than 95 %) of the Co layers of this film is ferromagnetic, as known from the previous Kerr effect measurements [15].

Table 1. Components of the dielectric permeability tensor for Co and Cu at the light wavelength of $\lambda$ = 632,8 nm.

| $\varepsilon_{xx}$(Co) | $\varepsilon_{xy}$(Co) | Q | Ref. |
|---|---|---|---|
| -11.5 - $i$18.32 | 0.89 - $i$0.89 | 0.055 - $i$0.013 | [34] |
| -12.0 - $i$19 | 0.56 - $i$0.08 | 0.023 + $i$0.01 | [35] |
| -8.19 - $i$16.38 | 0.499 - $i$0.1 | 0.027 +$i$0.007 | [36] |
| -12.5 - $i$18.46 | | | [37] |
| -11.5 + $i$18.31 | | | [38] |
| | | 0.043 + $i$0.007 | [39] |

| $\varepsilon_{xx}$(Cu) | | | Ref. |
|---|---|---|---|
| -11.64 - $i$1.64 | | | [33] |
| -11.6 - $i$1.84 | | | [40] |

As it is seen from Fig. 6 the method of effective medium describes satisfactory the observed monotonic increase of the $K(d_{Cu})$ function dependence. With taking into account the size dependence of the dielectric tensor components of ultrathin films on their thickness the consistency of the experiment data with the description will be improved [31]. Besides, reflection of light from the surface of metal ultrathin films changes significantly when their thickness is several atomic layers [41]. So, it is possible that multiple reflections of light from interfaces make a contribution to the observed monotonic increase of the Faraday rotation and this contribution rises with increasing of the Cu layer thickness.

### 3.2 Non-monotonic changes of the Faraday rotation

When the thickness of the copper layers is close to 0.9 nm and 1.8 nm, both the Faraday rotation angle and the magneto-optical coefficient decrease. The non-monotonic changes in the Faraday rotation with increasing of copper layers thickness in the multilayer Co/Cu films consist in decrease of the Faraday rotation angle and reducing of the magneto-optical coefficient $V$ at thicknesses of the copper layers close to 0.9 nm and 1.8 nm. The decrease of $\varphi$ can be caused by reduction in number of cobalt atoms bound in the ferromagnetic blocks, with increase of the "easy-plane" type anisotropy and with appearance of the antiferromagnetic exchange interaction between the ferromagnetic blocks of the nearest cobalt layers. The fact that these features are observed at the thicknesses of copper layers at which the exchange interaction of RKKY (Ruderman–Kittel–Kasuya–Yosida) between the cobalt layers is established requires consideration the possibility of influence of the size quantization of the electron density in the Cu layers on magnetic and magneto-optical properties of the multilayer films.

The antiferromagnetic exchange interaction between the ferromagnetic layers changes the magnitude of the coefficient $V$, but not the coefficient $K$. The antiferromagnetic exchange field prevents the alignment of the magnetic moments of the ferromagnetic cobalt layers to direction of the applied magnetic field as well as the demagnetization field. If the exchange energy and energy of demagnetization field is expressed as $\varepsilon_{AF} = \frac{1}{2}\gamma M_0^2 \cos 2\theta$ and $\varepsilon_{dem} = \frac{1}{2}N M_0^2 \cos^2\theta$, correspondingly, the magnetization along the field can be expressed as $M = \frac{H}{N+\gamma}$ and then $\varphi = md_{Co}K_{calc}\frac{H}{N+\gamma}$.

For the film with $d_{Cu} = 1.8$ nm, for which the deepest minimum in the $V(d_{Cu})$ dependence (open circle in Fig. 6) is observed and $V = 1.22 \times 10^2$ deg/(Oe cm) and $K_{calc} = 3.34 \times 10^2$ deg/(G cm), the exchange parameter $\gamma$ can be calculated as

$$\gamma = \frac{K_{calc} - 4\pi V}{V} = -7.97, \qquad (7)$$

Assuming that the magnetization of the adjacent antiferromagnetic coupled FM Co layers $M_0$ is equal to 495 G, the exchange field is

$$H_{ex}^{AF} = \frac{|\gamma|M_0}{2} \approx 2 \, kOe$$

and the energy of this AFM exchange coupling is equal to

$$J = \frac{|\gamma|M_0^2 d_{Co}}{2} = 0.08 \, \frac{erg}{cm^2}.$$

This value of the exchange coupling energy is close to those obtained for the $[Co/Cu(111)]_m/Co$ systems in different experiments: 0.1 erg/cm$^2$ (Ref. [42]), 0.06 erg/cm$^2$ (Ref. [43]), and 0.05 erg/cm$^2$ (Ref. [44]).

However, the AF exchange interaction cannot explain the decrease in the Faraday rotation angle of the films with $d_{Cu}$ near 0.9 and 1.8 nm in the fields larger than the flip field ($2H_{ex}^{AF}$) of the magnetic moments of the FM layers. This decrease in $\varphi$ can be caused by magnetization reduction of the whole volume of the ferromagnetic blocks in the films. The Co/Cu films under study contain some number of cobalt ions that are not included in ferromagnetic blocks, but are combined into clusters whose blocking temperature is lower than the room temperature. Note that the "loss" of cobalt atoms by ferromagnetic blocks entails decrease in magnetization and magnitude of the Faraday rotation, but it alone does not lead to decrease in the magneto-optical coefficients $K$ and $V$, since the magnetic susceptibility remains equal to $1/N$.

To determine the number of these "loose spins" depending on the thickness of the copper layers, the Faraday rotation angles in the state of magnetic saturation of all films were determined. The applied magnetic field was not strong enough to reach the saturated state in most of the films. Therefore, the $\varphi(H)$ dependences for each film were converted in functions of the internal magnetic field

$$H_{int} = H - 4\pi\frac{\varphi(H)}{K_{calc}md_{Co}}$$

and the Faraday angle in saturation, $\varphi_{sat}$, was determined by the linear extrapolation of the $\varphi(H_{int})/H_{int}$ plots versus $\varphi$ to zero. The magnetizations of the films in saturation were defined as

$$M_{sat}(d_{Cu}) = \frac{\varphi_{sat}(d_{Cu})}{K_{calc}(d_{Cu})md_{Co}}.$$

As an example Fig. 7 illustrates the procedure of determination of $\varphi_{sat}$ and $M_{sat}$ for particular nanofilms.

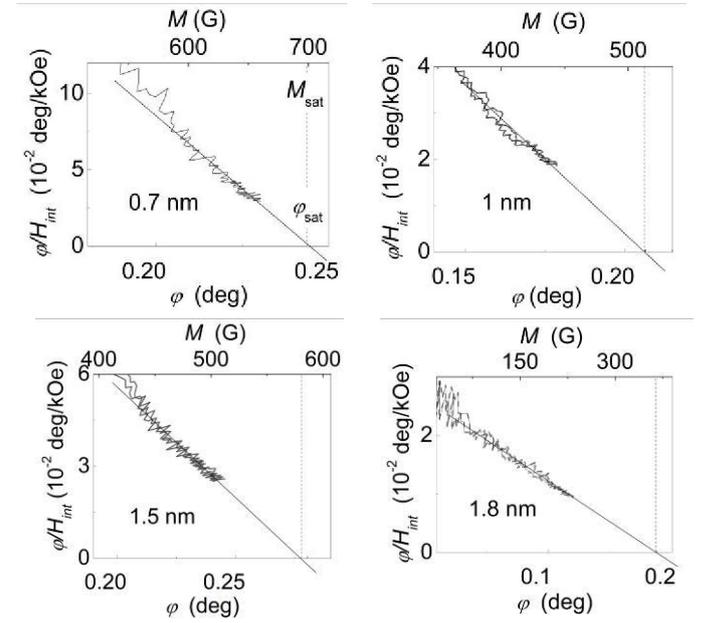

Fig. 7. Determination of the Faraday rotation angle $\varphi_{sat}$ and magnetization $M_{sat}$ of the Co/Cu multilayers at the saturated magnetic states for the samples with the Cu layer thicknesses $d_{Cu} = 0.7$, 1.0, 1.5, and 1.8 nm. The top horizontal axis for magnetization is $M = \frac{\varphi}{K_{calc}(d_{Cu})md_{Co}}$.

The obtained values of $M_{sat}$ correspond to the magnetization of saturation of ferromagnetic blocks and sufficiently large superparamagnetic clusters. Contribution from magnetization of the "loose spins" (up to 20 atoms per cluster) [12,45,46] is negligible in the fields of order 10 kOe at room temperature.

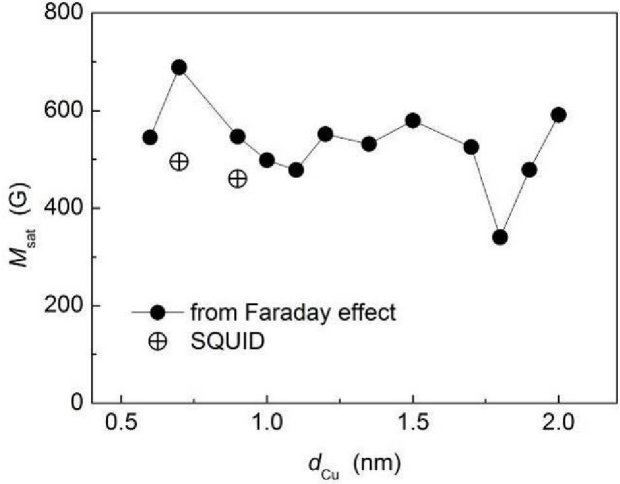

Fig. 8. Dependences of the saturation magnetization $M_{sat}$ on the Cu layer thickness for the $[Co/Cu(d_{Cu})]_{20}$ nanofilms. The magnetization values obtained from the Faraday rotation angle dependences $\varphi(H)$ are shown as solid circles, while the values obtained from magnetic SQUID measurements (for films with $d_{Cu}$ = 0.7 nm and 0.9 nm) are represented by crossed circles.

It can be seen from Fig. 8 that the magnitude $M_{sat}$ depends on the thickness of the copper layers in a non-monotonic way and demonstrates minima near $d_{Cu}$ = 1.0 - 1.1 nm and 1.8 nm. Such decrease can indicate a fine fragmentation of the cobalt layers in the same films, namely an increase in the number of the "loose spins". The $M_{sat}$ values measured in the fields up to 40 kOe using a SQUID-magnetometer for two films with $d_{Cu}$ = 0.7 nm and 0.9 nm (crossed circles in Fig. 8) are presented as well. The fact that the saturation magnetization values of these two films measured directly are close to the values determined from the magneto-optical measurements of the Faraday rotation confirms the reliability of the latter.

It should also be noted that the anisotropy of magnetic defects of type "easy axes in the plane" of the film can make a significant contribution to the non-monotonic changes in the coefficient $V$ of the film depending on thickness of the Cu layers. This anisotropy contribution rises in the films with increase of fragmentation degree of the ferromagnetic disks and roughness of the Co/Cu interfaces in the multilayer pillars.

Quasi-one-dimensional interface defects and edge magnetic defects form a net of local magnetic anisotropy of the "easy axes in the plane" type, the random directions of the easy axes of which are close to the plane of the film. This inhomogeneous anisotropy prevents the alignment of magnetic moments along the field and can lead to a decrease in the coefficient $V$ according to the expression

$$V = \frac{K}{4\pi + \gamma + \kappa_{loc} v}.$$

Here $\kappa_{loc}$ and $v$ are integral effective constants which characterize the local anisotropy magnitude of the "effective easy plane" type and the fraction of the volume where it takes place, respectively.

Thus the non-monotonic changes of the magneto-optical coefficient $V$ in the Fig. 6 are due to influence of both antiferromagnetic exchange interaction and the local magnetic anisotropy of the films on process of magnetization. In any case, the influence of both contributions is observed for the thicknesses of copper layers at which size quantization of the electron density occurs in the layers.

### 3.3 Surface morphology of the nanofilms

It is known that the antiferromagnetic exchange coupling between FM layers in FM/NM multilayers is induced by the quantum size effect in the NM layers, namely a redistribution of the electron density in the volume of the NM layers. This redistribution induces electric fields, which affect on deposition of the FM atoms during formation of the FM layers in the multilayer structures. Such electron distribution can induce structural changes of deposited metal layers [47-50] and impact on distance between atomic layers [51-53], quantum dots distribution and structure of FM clusters on surface of non-ferromagnetic metal [54,55]. Therefore, it is most likely that such spatial redistribution of electron density in the volume of the Cu layers affect also on formation of the Co layers und appearance of the fine fragmentation of the antiferromagnetically exchange coupled Co layers in the $[Co/Cu(111)]_{20}$ nanofilms.

Morphological features of metal films with a thickness of several atomic layers are determined by the substrate, interface and the competition mainly between the energy anisotropy, the surface stress and the energy associated with this spatial quantization of electron density. It can be expected that the fragmentation of the cobalt layers, caused by this redistribution of the electron density in the films, will affect the surface structure of the whole film.

Surfaces of the films with $d_{Cu}$ = 0.9, 1.5 and 1.8 nm were studied using Atomic Force Microscopy. Figure 9 shows the distribution of electric potential on the surfaces of these three films. The surface potential of the films, in which the copper layers undergo spatial electronic quantization ($d_{Cu}$ = 0.9 and 1.8 nm) indicates a much larger number of surface defects than the film with $d_{Cu}$ = 1.5 nm, in which no electronic spatial quantization occurs.

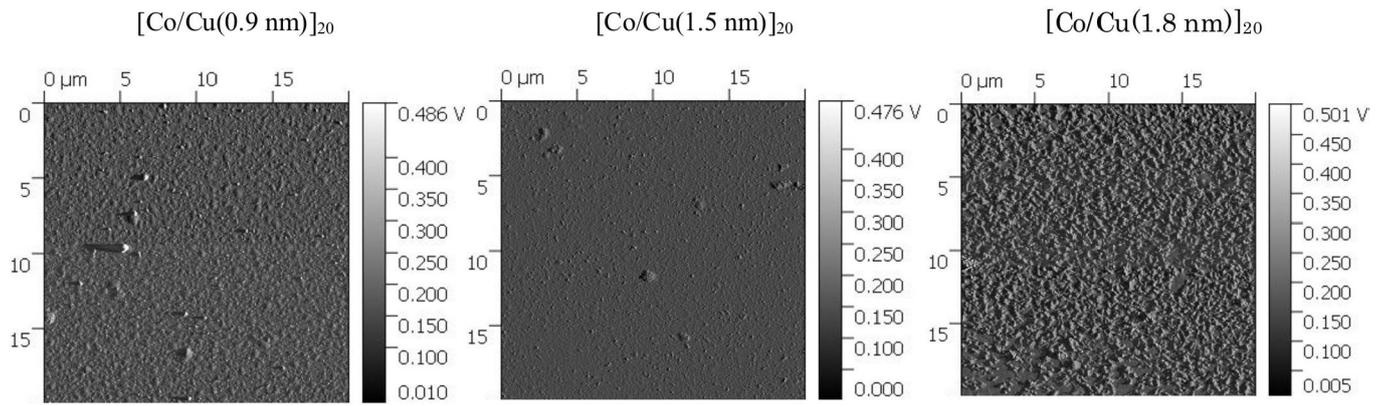

Fig. 9. The images of surface potential of the nanofilms with $d_{Cu}$ = 0.9, 1.5, and 1.8 nm, obtained using Atomic Force Microscopy. The contrast axis displays roughly the height of the surface defects.

## 3. Conclusions

Magnetic field induced changes in the Faraday effect in periodic [Co(0.8 nm)/Cu(111)($d_{Cu}$)]$_{20}$ nanofilms having the same thickness of all cobalt layers and the thickness of the copper layers is varied from 0.6 to 2 nm, were studied. Monotonic rise of the $\varphi(H)$ dependence with increasing of the Cu layer thickness and two minima on background of this rise were detected. It was demonstrated that the monotonic increase of the Faraday rotation angle is well described within the approximation of an effective optical medium. The non-monotonic variations in the $\varphi(d_{Cu})$ dependence, which were observed as two minima of the Faraday rotation angle at $d_{Cu}$ = 1.8 and 1.0 nm are associated with decrease of magnetization due to the "loose spins" as result of the Co layer's fragmentation. We suggest that increasing of the magnetic fragmentation in these films appears during deposition of the Co layers in conditions of the quantum size effect in the Cu layers. Changes in the surface defects of the nanofilms, observed using Atomic Force Microscopy, confirm indirect influence of the electronic quantization in the Cu layers on the defect structure of the multilayer [Co/Cu(111)]$_m$ nanofilms.

## Acknowledgments

The authors would like to thank V. A. Pashchenko for discussions and helpful comments. The research performed in the University of Aveiro was developed within the scope of the project CICECO-Aveiro Institute of Materials, UIDB/50011/2020 & UIDP/50011/2020, financed by national funds through the FCT/MEC and when appropriate co-financed by FEDER under the PT2020 Partnership Agreement.

## References

[1] C. Rizal, B. Moa and B.B. Niraula, Magnetochemistry 2 (2016) 22.
[2] M.M. Krupa, Adva. Eng. Tech. 2 (2017) 1.
[3] C.O. Avci, M. Mann, A.J. Tan, P. Garbardella, G.S.D. Beach, Appl. Phys. Lett. 110 (2017) 203506.
[4] G. Antarnusa, P.E. Swastika, E. Suharyadi, Journal of Physics: Conf. Series 1011 (2018) 012161.
[5] D. F. R. Vasquez, V. Biziere, B. Warot-Fonrose, T. Wade, C. Gatel, Nano Lett. 16 (2016) 1230.
[6] P. Zahn, I. Mertig, Phys Rev B 63 (2001) 104412.
[7] I. Ennen, D. Kappe, T. Rempel, C. Glenske, A. Hütten, Sensors 16 (2016) 904.
[8] E.M. Shima, L. Salamanca-Riba, R.D. McMichael, T.P. Moffat, J. Electrochem. Soc. 149 (2002) C439.
[9] P. Vavassori, F. Spizzo, E. Angeli, D. Bisero, F. Ronconi, J. Magn. Magn. Mater 262 (2003) 120.
[10] F. Spizzo, E. Angeli, D. Bisero, P. Vavassori, F. Ronconi, J. Magn. Magn. Mater 242-245 (2002) 473.
[11] N. Rajasekaran, J. Mani, B,G. Tóth, G. Molnár, S. Mohan, I. Bakonyi, J. Electrochem. Soc. 162 (2015) D204.
[12] I. Bakonyi, L. Péter, Z. Rolik, K. Kiss-Szabo, Z. Kupay, J. Tóth, L.F. Kiss, J. Pádár, Phys. Rev. B 70 (2004) 054427.
[13] J. Balogh, M. Csontos, D. Keptás, G. Mihály, Solid State Commun. 126 (2003) 427.
[14] I.N. Lukienko, N.F. Kharchenko, V.M. Khrustalev, V.N. Savytskiy, A.V. Fedorchenko, V.A. Desnenko, A.N. Stetsenko, V.V. Zorchenko, Low Temp. Phys. 38 (2012) 848.
[15] I.N. Lukienko, N.F. Kharchenko, A.N. Stetsenko, J. Nano- and Electronic Physics 10 (2018) 06041.
[16] V.V. Zorchenko, A.N. Stezenko, A.G. Anders, K.V. Kutko, Low Temp. Phys. 31 (2005) 505.
[17] R. Hammerling, C. Uiberacker, J. Zabloudil, P. Weinberger, L. Szunyogh, J. Kirschner, Phys. Rev. B 66 (2002) 052402.
[18] E. H. Brandt, Low Temp. Phys. 27 (2001) 723.
[19] J. Dubowik, Phys. Rev. B 54 (1996) 1088.
[20] L. Chen, S. Wang, R.D. Kirby in *Handbook of Advanced Magnetic Materials: Properties and Applications*, edited by D. J. Sellmyer, Y. Liu, D. Shindj (Tsinghua University Press, Springer , 2005) Vol. IV, p. 241.
[21] V.N. Antonov, L. Uba, S. Uba, A.N. Yaresko, A.Ya. Perlov, V.V. Nemoshkalenko, Low Temp. Phys. 27 (2001) 425.
[22] Y.B. Xu, O.Y. Jin, Y. Zhai, Y.Z. Miao, M. Lu, H.R. Zhai, J. Magn. Magn. Mater. 126 (1993) 541.
[23] T. Katayama, Y. Suzuki, M. Hayashi, A. Thiaville, J. Magn. Magn. Mater 126 (1993) 527.
[24] R. Megy, A. Bounouh, Y. Suzuki, P. Beauvillain, P. Bruno, C. Chappert, B. Lecuyer, P. Veillet, Phys. Rev. B 51 (1995) 527.
[25] R. Atkinson, G. Didrichsen, W.R. Hendren, I.W. Salter, R.J. Pollard, Phys. Rev. B 62 (2000) 12294.
[26] Š. Višňovský, K. Postava, T. Yamaguchi, Optics Express 9 (2001) 158.
[27] P. Yeh, Surface Science 96 (1980) 41.
[28] J. Zak, E.R Moog, C. Liu, S.D. Bader, J. Magn. Magn. Mater 89 (1990) 107.
[29] R. Atkinson, J. Magn. Magn. Mater 115 (1992) 353.
[30] M. Abe, M. Gomi, Jpn. J. Appl. Phys. 23 (1984) 1580.
[31] R. Atkinson, P.M. Dodd, J. Magn. Magn. Mater. 173 (1997) 202.
[32] K. Shinagawa, in *Magneto-Optics*, edited by S. Sugano and N. Kojima (Springer, Berlin, 1999), pp.137-177.
[33] D.W. Lynch, W.R. Hunter in *Handbook of Optical Constants of Solids,* edited by E. D. Palik (Orlando, Academic Press, 1985), p. 275.
[34] R.M. Osgood, K.T. Riggs, A. E. Johnson, J. E. Mattson, C. H. Sowers, S. D. Bader, Phys. Rev. B 56 (1997) 2627.
[35] G.S. Krinchik, Izv. Acad. Nauk SSSR Ser. Fiz. 21 (1964) 1293.
[36] G.A. Prinz, J.J. Krebs, D.W. Forester, W.G. Maisch, J. Magn. Magn. Mater. 15-18 (1980) 779.
[37] P.B. Johnson, R.W. Christy, Phys. Rev. B 9 (1974) 5056.
[38] J.H. Weaver in CRC *Handbook of Chemistry and Physics,* (Ed. R.C. Weast, M.J. Astly and W.H. Beyer) (CRC Press, Boca Raton, 1988) p. E-387ff.
[39] Z.Q. Qiu, J. Pearson, S.D. Bader, Phys. Rev. B 46 (1992) 8195.
[40] P.B. Johnson, R.W. Christy, Phys. Rev. B 6 (1972) 4370.

[41] A. E. Kaplan, Radio Eng. Electron, Phys. 9 (1964) 1476; A.E. Kaplan, J. Opt. Soc. Am. B 35, (2018) 1328.
[42] S.S.P. Parkin, R. Bhadra, K.P. Roche, Phys Rev Lett 66 (1991) 2152.
[43] A.S. Samardak, P.V. Kharitonskii, Yu.D. Vorob'ev, L.A. Chebotkevich. Phys. Met. Metallogr. 98 (2004) 360.
[44] D.H. Mosca, F. Petroff, A. Fert, P.A. Schroeder, W.P. Pratt, R. Laloee, J. Magn. Magn. Mater. 94 (1991) L1.
[45] J.C. Slonczewski, J. Appl. Phys. 73 (1993) 5957.
[46] J.J. de Vries, G.J. Strijkers, M.T. Johnson, A. Reinders, W.J.M. de Jonge, J. Magn. Magn. Mater. 148 (1995) 187.
[47] L. Floreano, D. Cvetko, F. Bruno, G. Bavdek, A. Cossaro, R. Gotter, A. Verdini, A. Morgante, Prog. Surf. Science 72 (2003) 135.
[48] M.C. Tringides, M. Jalochowski, E. Bauer, Phys. Today 60 (2007) 50.
[49] Y. Jia, M.M. Özer, H.H. Weitering, Z. Zhang in *Nanophenomena at surfaces*, edited by M. Michailov (Springer series in surface sciences, Springer, Berlin, Heidelberg 2011) Vol. 47, pp. 67-112.
[50] D.L. Price, V. Henner, M. Khenner, J. Appl. Phys. 124 (2018) 174302.
[51] P. Czoschke, H. Hong, L. Basile, T.-C. Chiang, Phys. Rev. B 72 (2005) 035305.
[52] Y. Jiang, K. Wu, Zh. Tang, Ph. Ebert, E.G. Wang, Phys. Rev. B 76 (2007) 035409.
[53] T.-C. Chiang, Chinese J. Phys. 43 (2005) 154.
[54] H. Hong, L. Basile, P. Czoschke, A. Gray, T.-C. Chiang, Appl. Phys. Lett. 90 (2007) 051911.
[55] S. Pick, V.S. Stepanyuk, A.L. Klavsyuk, L. Niebergall, W. Hergert, J. Kirschner, P. Bruno, Phys. Rev. B 70 (2004) 224419.